# Characterizing the Influence of Charge Extraction Layers on the Performance of Triple Cation Perovskite Solar Cells

*Johanna Siekmann, Ashish Kulkarni,* Samah Akel, Benjamin Klingebiel, Michael Saliba, Uwe Rau, and Thomas Kirchartz**

J.S. and A.K. contributed equally to this work.

*Johanna Siekmann, Samah Akel, Benjamin Klingebiel, Uwe Rau, Thomas Kirchartz*
IEK5-Photovoltaik, IEK5-Photovoltaik, Forschungszentrum Jülich GmbH, Wilhelm-Johnen-Straße, 52428 Jülich, Germany.

*Ashish Kulkarni, Michael Saliba*
Helmholtz Young Investigator Group FRONTRUNNER, IEK5-Photovoltaik, Forschungszentrum Jülich, 52425 Jülich, Germany

*Michael Saliba*
Institute for Photovoltaics, University of Stuttgart, Pfaffenwaldring 47, 70569 Suttgart, Germany

*Thomas Kirchartz*
Faculty of Engineering and CENIDE, University of Duisburg-Essen, Carl-Benz-Str. 199, 47057 Duisburg, Germany

Corresponding Authors
E-mail: a.kulkarni@fz-juelich.de, t.kirchartz@fz-juelich.de

## Abstract

Selecting suitable charge transport layers and suppressing non-radiative recombination at interfaces to the absorber layer are vital to maximize the efficiency of halide perovskite solar cells. In this work, high quality perovskite thin films and devices are fabricated with different fullerene-based electron transport layers and different self-assembled monolayers as hole transport layers. We then perform a comparative study of a significant variety of different electrical, optical and photoemission-based characterization techniques to quantify the properties of the solar cells, the individual layers and importantly the interfaces between them. In addition, we highlight the limitations and problems of the different measurements, the insights gained by combining different methods and the different strategies to extract information from the experimental raw data.

## 1. Introduction

Electron and hole transport layers for lead-halide perovskite solar cells have to fulfil several requirements simultaneously:[1] (a) provide low recombination at its interfaces to the absorber layer (passivation),[2-4] (b) provide the appropriate band energy alignments to the absorber layer to allow efficient extraction of one type of charge carrier while blocking the other one[5, 6] (selectivity) and, finally, (c) to minimize optical losses (parasitic light absorption) and resistive losses. Especially, in inverted or *p-i-n* structured perovskite solar cells, choosing the best charge transport layers (CTLs) is challenging as there are a multitude of different candidates to choose from that include a variety of self-assembled monolayers (SAM)[7, 8], polymers or fullerene derivatives[9]. To avoid having to choose randomly and to perform a trial-and-error based optimization, gaining insights into the effect of charge-transport layers on the performance, recombination and charge collection in perovskite solar cells is of critical importance.[10-12] There is a large amount of literature dealing with perovskite-CTL interfaces showing that the modification of perovskite surfaces[5, 6, 13, 14] as well as the choice of suitable transport layers frequently has a considerable influence on device performance.[2, 7, 8, 15, 16] However, characterization of the interfaces and interlayers is very often done by using a set of individual methods whose results are presented separately and often without any quantitative and model-based analysis. What is typically missing are experimentally determined parameters of the interface or the CTL that are quantitatively consistent with solar cell performance. Furthermore, there is typically a lack of comparisons between different measurement methods as well as consistency checks that critically examine the experimentally obtained parameters. The consequence of this situation is that trends are reported while the community is still quite far from being able to do quantitative characterization that could be validated with a model of the interfaces and the device.

Here, we apply a set of state-of-the-art characterization methods to triple-cation perovskite samples employing a range of different electron and hole transport layers. The methods used include electrical characterization, optical spectroscopy and surface physical methods based on photoemission spectroscopy that we apply to a variety of samples ranging from single layers to full devices. The aim of the present work is twofold: First we attempt to identify the efficiency limiting mechanisms in the different device stacks and identify the strengths and weaknesses of the different electron transport layers (ETLs) and hole transport layers (HTLs) in combination with our specific perovskite layer. Secondly, we want to study the problem from the point of view of method development and method comparison. Hence, we focus on the problems or challenges in the analysis of the different characterization methods as well as on insights that can be obtained from analyzing different experiments in combination.

A crucial requirement for identifying suitable ETLs and HTLs for a given perovskite absorber layer is the band alignment at the interfaces towards the transport layers as well as the resulting band diagram.

The band alignment is often determined by ultraviolet photoelectron spectroscopy (UPS).[17] The correct quantification of the valence band off-set for perovskite is still an ongoing discussion.[18-20] In this work we show the difference between three methods for the determination of the valence band edge and discuss the relation between the observed energy levels and the predicted performance of a device using numerical simulations of current-voltage curves. To complement and compare the UPS data, we use space charge limited current (SCLC) measurements on single carrier devices to gain insights on the band alignment between perovskite and ETL. We find that both the band offset and the electron mobility in the transport layer change the ratio between the forward and reverse current density.

In addition to a suitable band offset, the transport layers should also feature low series resistances and minimize losses due to non-radiative recombination in the cell. We determined the series resistance by comparing dark and illuminated current voltage curves of the full devices.[12, 21, 22] Poor CTLs can minimize the fill factor (*FF*) of a cell. This can be due to a high series resistance or an ideality factor $n_{id}$ > 1. We quantify both types of losses for solar cells with varying HTLs and ETLs and find that the loss in *FF* due to a high series resistance can occur due to poor HTL and ETL. The loss due to non-ideal diode behavior appears to be not as strongly influenced by the transport layers.

While there are relatively simple approaches to quantify recombination losses at open circuit in layers, layer stacks and full devices, it is currently quite difficult to study the effect of transport layers on collection losses as well as resistive effects. Furthermore, gaining deep insights into recombination dynamics at interfaces is highly challenging due to the superposition of different effects that modify transient or frequency domain experiments of layer stacks or full devices. Here, we overcome these challenges by combining several measurement methods. First we use the combination of voltage dependent photoluminescence (PL)[23-26] and transient PL[27] that can give information about the charge carrier extraction. In the case of transient PL with contact layers, there is no clear way to tell whether a fast decay time is due to good extraction or high recombination.[27] However, a consistency check with the voltage-dependent PL shows that the fast decay is due to good extraction. Finally, we investigate non-radiative recombination by combining steady-state PL and transient PL. Through the comparison of the Fermi level splitting at one Sun intensity we can show that the long lifetime of the transient PL results from a de-trapping of charges that takes place at long times.

Our results highlight the importance of critical analysis of different characterization methods that are used to gain insights into the properties of CTLs. By showing a broad range of characterization methods we give the community tools to select the fitting CTL for their perovskite instead of testing various ETLs and HTLs.

# 2. Results

To investigate the influence of different charge transport layers on the performance of perovskite solar cells, we employ four different HTLs and three different ETLs. For HTLs, poly[bis(4-phenyl)(2,4,6-trimethylphenyl)amine] (PTAA), and carbazole based self-assembled monolayers (SAMs) with phosphonic acid anchoring group such as 2-PACz, MeO-2PACz, and Me-4PACz were deposited while $C_{60}$ fullerene, and other fullerene derivatives such as PCBM and CMC were deposited as ETLs. A triple cation composition ($Cs_{0.05}(FA_{0.83}MA_{0.17})_{0.95}Pb(I_{0.83}Br_{0.17})_3$) with a bandgap ($E_g$) of 1.62 eV was employed in our study which was deposited by improving the perovskite ink-substrate interaction using triple (DMF:DMSO:NMP) co-solvent system, sandwiched in between the HTL and ETL. It is important to note that the aim of the present work is not to employ different combinations of HTLs and ETLs and report the best device efficiency, but rather to investigate the influence that different electron and hole transport layers have on the experimentally accessible properties of interfaces, of recombination, of charge extraction and of device performance. In order to somewhat limit the range of options, the $C_{60}$ and Me-4PACz were maintained as default ETL and HTL respectively, while the respective other transport layers were varied. The $C_{60}$ and Me-4PACz were maintained as default because they have been reported to demonstrate highest efficiency in wide $E_g$ single and silicon-perovskite tandem junction devices.[16]

## 2.1 Energy level alignment

The alignment of energy levels, i.e. conduction band edges of perovskite and ETLs as well as the valence band edges of perovskite and HTLs have an influence on both recombination and extraction. We could define a band offset e.g. at the perovskite-ETL interface as $\Delta E_c = \chi_{\mathrm{ETL}} - \chi_{\mathrm{pero}}$, where $\chi$ represents the electron affinity of ETL ($\chi_{\mathrm{ETL}}$) and perovskite layer ($\chi_{\mathrm{pero}}$). The optimum offset $\Delta E_c$ is not necessarily the same for these two requirements. Given that interfaces break lattice periodicity and hence increase the likelihood of defects, device design in photovoltaics in general will always aim at minimizing the density of electrons and holes that face each other across an interface. For a given voltage, these densities will increase when the interfacial band gap is lowered or the conduction band offset $\Delta E_c$ is increased. Thus, to minimize recombination, the interfacial band gap $E_{\mathrm{int}} = E_g - \Delta E_c$ at the perovskite-ETL interface should ideally be identical or even higher than the bulk band gap $E_g$ of the absorber, implying that the band offset would be zero or even positive (forming a barrier for extraction).[28-30] In contrast, for efficient charge extraction at short circuit, a higher value of $\Delta E_c$ would be acceptable[30] as long as $\Delta E_c$ does not become negative, i.e. there is no barrier for electron extraction. The typical method to study the alignment of energy levels is ultraviolet photoelectron spectroscopy (UPS) which can be performed on different layer stacks. UPS directly provides the work function and the valence band edge of the surface of the measured layer. However, especially the determination of the valence band edge is plagued by the difficulty to precisely assign the band edge to a continuously changing spectrum. The electron affinity cannot directly be determined from UPS but can be obtained by adding the optical band gap of the material to the valence band edge. Thus, it faces the same or even more uncertainties as the valence band edge.

**Figure 1a** shows the conduction band minimum (CBM), the Fermi energy and the valence band maxima (VBM) at vacuum energy $E_{vac} = 0$ eV for different possible components of a *p-i-n* solar cell. For both different HTLs and the different ETLs, the valence band edge can be determined by a linear fit as explained in the Supporting Information. For the conduction band edge, band gaps from the literature are assumed. For the HTLs, i.e. PTAA and SAMs, the following band gaps are assumed. The band gaps of $E_g = 3.4$ eV (PTAA), $E_g = 3.2$ eV (MeO-2PACz), $E_g = 3.4$ eV (2PACz)[8] and of $E_g = 3.3$ eV (Me-4PACz) were adopted form references [8, 16]. For the HTL, the ionization energy (represented by $E_i = E_{vac} - E_V$) is the decisive parameter for the carrier transport from perovskite to HTL. PTAA shows the lowest value of $E_i$ while 2PACz showed the highest $E_i$. For the electron transport, the electron affinity of the ETL is important. For all three ETLs a band gap of $E_g = 2.3$ eV is assumed, resulting in $C_{60}$ having the largest electron affinity and CMC having a slightly smaller affinity than PCBM.

Decisive for the band diagram are of course also the band positions of the perovskite. In contrast to the CTLs, there is no conclusive agreement on the optimal evaluation of the measurement data for the determination of the valence band edge as per the UPS. Three different approaches are proposed in the literature. The linear evaluation, analogous to the evaluation of the CTL, the evaluation via a logarithmic scale[19] and the fitting of a Gaussian function[18] to the valence band edge. The three methods are shown in **Figure S1**. **Figure 1b** shows the simulated[31] band diagram and **Figure 1c** the current-density – voltage (*JV*) curve of a cell with Me-4PACz as HTL and $C_{60}$ as ETL, for all three evaluation methods. First, we consider the linear evaluation. Since the linear evaluation is not sensitive to the background, the measurements can be compared well with each other, implying that it is well suited to study trends. However, due to the fact that the valence band edge in perovskite is not very steep, the linear method leads to an overestimation of the ionization energy.[19] This results in a significant electron extraction barrier of $\Delta E_c = \chi_{ETL} - \chi_{pero} = -62$meV between the conduction band of the perovskite and $C_{60}$ for the band diagram. For the other fullerenes this barrier is even higher (Figure S4a). The Me-4PACz/perovskite interface shows a band offset of $\Delta E_V = E_i^{pero} - E_i^{HTL} = 170$meV. The simulated *JV*-curve for the linear evaluation method and $C_{60}$ as ETL shows an S-shape and a poor *FF*. As a second method, we consider the valence band edge on a semi-logarithmic scale and define the intersection between the background and the valence band edge as in the linear method. On the logarithmic axis, the result is much more dependent on the background than with the linear method. However, since the background depends on the measurement conditions, the comparability between different measurements analyzed with the logarithmic method is not as good as with the linear evaluation. However, the ionization energies are smaller and do not result in electron extraction barriers in the band diagram and lead to more realistic *JV* simulations with an open-circuit voltage $V_{OC}$ = 0.92 V and *FF* = 76%. The HTL/perovskite interface shows an extraction barrier for holes of $\Delta E_V = -330\text{meV}$. We enable tunneling through thin barriers in the drift-diffusion simulation as we assume that this is a realistic transport scenario for the monolayer-based HTLs. This assumption causes only an increase in $V_{OC}$ for the simulated *JV*-curve of Me-4PACz to MeO-2PACz. Although there is a hole barrier for Me-4PACz ($\Delta E_V = -330\text{meV}$) and just an offset of $\Delta E_V = 70\text{meV}$ for MeO-2PACz. Despite tunneling, the barrier height seems to prevent the simulation for 2PACz from converging for higher voltages. Further description of the band diagrams and *JV*-curves are given in chapter 2.1.1 and 2.1.2 of the SI. As an alternative to both methods, a Gaussian function can be fitted to the leading edge of the UPS spectra and the VBM position is determined only by the position and standard deviation sigma of the Gaussian function. In contrast to the two aforementioned methods the problematic intersection with the background level (which is influenced by the measurement conditions) is not used. However, with the Gaussian fit the problem remains where to define the exact position of the VBM. Endres et al. [18] compared DFT simulations with experimental UPS and IPES data and suggested a correction factor of $2.9 \times \sigma$ if the Gaussian fit is used, where $\sigma$ is the standard derivation of the Gaussian. The last method results in an ionization energy that is in between the linear and logarithmic methods and is not sensitive to the background. The simulated *JV*-curve with $V_{OC}$ = 1.10V and *FF*=78% fits best to the measured samples. In conclusion, we note that the translation of photoemission spectroscopy data into an approximately realistic device model strongly depends on the choice of energy levels even if we assume that the problematic interface is only towards the ETL and the monolayer-based HTLs can extract holes by tunneling through a barrier.

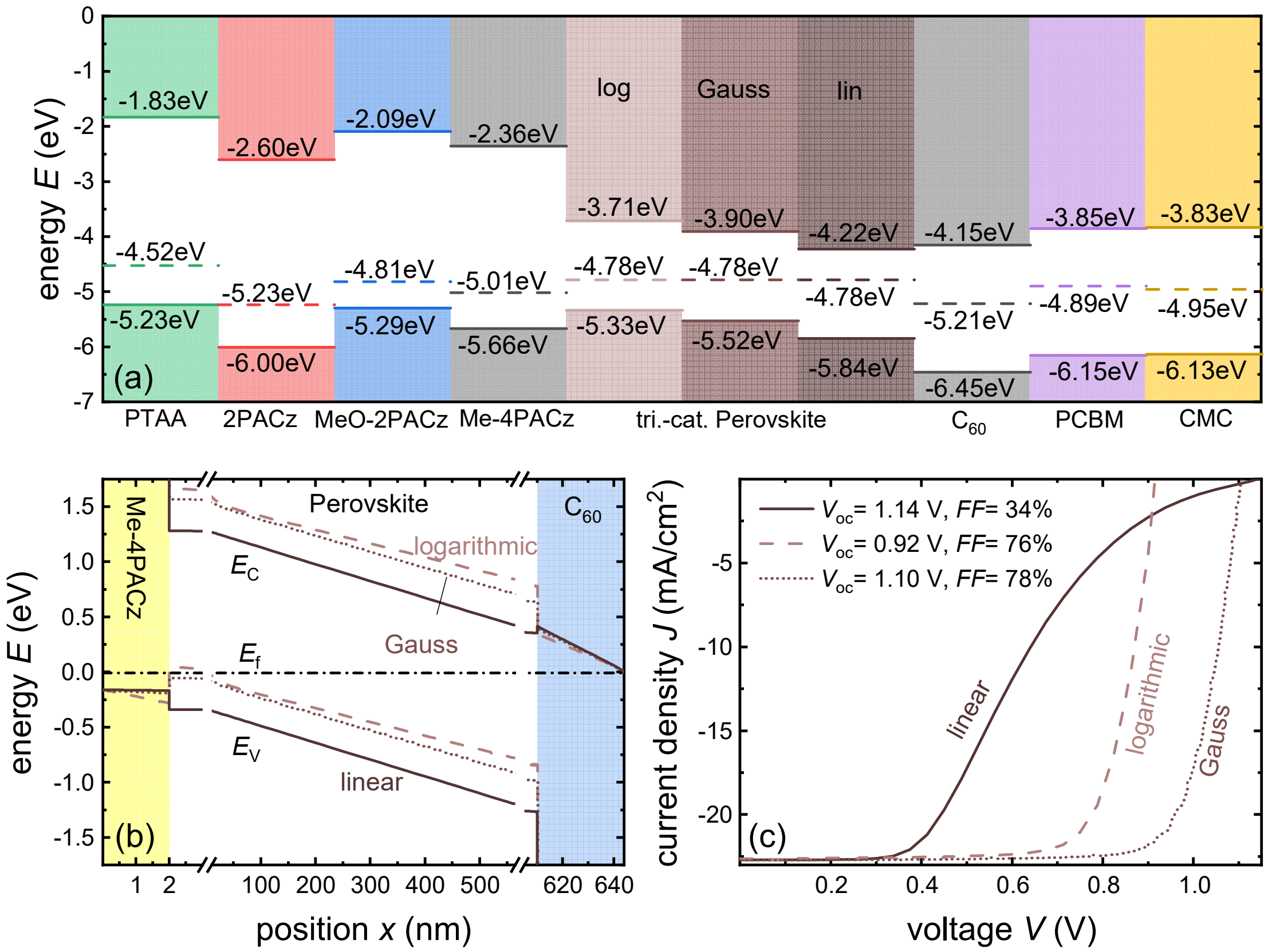

**Figure 1:** (a) Schematic of the energy-levels for different HTLs, perovskite on Me-4PACz and different ETLs deposited on top of the perovskite. We evaluate the UPS measurement in three different ways, labelled logarithmic, Gauss and linear. Each method leads to a different valence band offset. b) Simulated band diagram of a cell with Me-4PACz as HTL, $C_{60}$ as ETL. The three different UPS methods lead to different possible band offsets between the CTLs and the perovskite. c) The current voltage curve for the three different band diagrams; the linear method leads to a simulated $V_{OC}$ = 1.14V but a s-shape (*FF*=34%) *JV*-curve, for the logarithmic method the $V_{OC}$ = 0.92V and *FF*=76% and the Gauss method leads to the best cell with $V_{OC}$ = 1.10V and *FF* = 78%.

## 2.2 Single carrier devices

Measuring *JV* -curves of single-carrier devices is a highly popular method in perovskite photovoltaics that is regularly used to either determine mobilities[32, 33] or trap densities[34, 35]. Furthermore, it could be used in general to identify differences in injection barriers[36] although the method has so far been rarely applied. In single carrier devices, recombination is of no relevance as only transport of majority carriers is required to obtain current flow. This transport of majority carriers is non-ohmic as long as the semiconductor is sufficiently intrinsic to avoid the dark conductivity of the semiconductor to dominate transport. In the absence of any considerable dark conductivity but in the presence of two contacts that efficiently inject and extract one type of charge carriers, it is the mobility of the semiconductor, the injection barrier and the amount of space charge that dominate and limit the amount of current that can flow through the semiconductor at a given voltage. In the absence of contact- or defect-related effects, the space charge of injected carriers limits the current flow, and one observes space-charge limited conduction, which allows the detection of mobilities[32, 37]. If the space charge of charged defects has a significant impact on the total space charge, one may observe the so-called trap-limited current[34, 38] that allows the determination of defect densities. In halide perovskites, however, neither situation is easily achieved. Instead, the current is typically heavily affected by space charge induced by mobile ions[39, 40] – i.e. a special case of charged defects – and by the injection barriers at either contact. Thus, the analysis of single carrier *JV* -curves requires a critical assessment of the assumptions of the used analysis approaches. Important insights are obtained from the comparison of all four branches of the *JV*-curves[41], i.e. to compare both the scanning direction (high to low voltages or vice versa) and the polarity (e.g. injection of electrons from one contact or from the other).

We fabricated electron-only devices by deposition of $SnO_2$ on ITO with atomic layer deposition followed by spin coating the perovskite and evaporation of $C_{60}$ fullerene. BCP and Ag were deposited as back contacts. The devices were measured in the dark from -3 to 3V and from 3 to -3V.

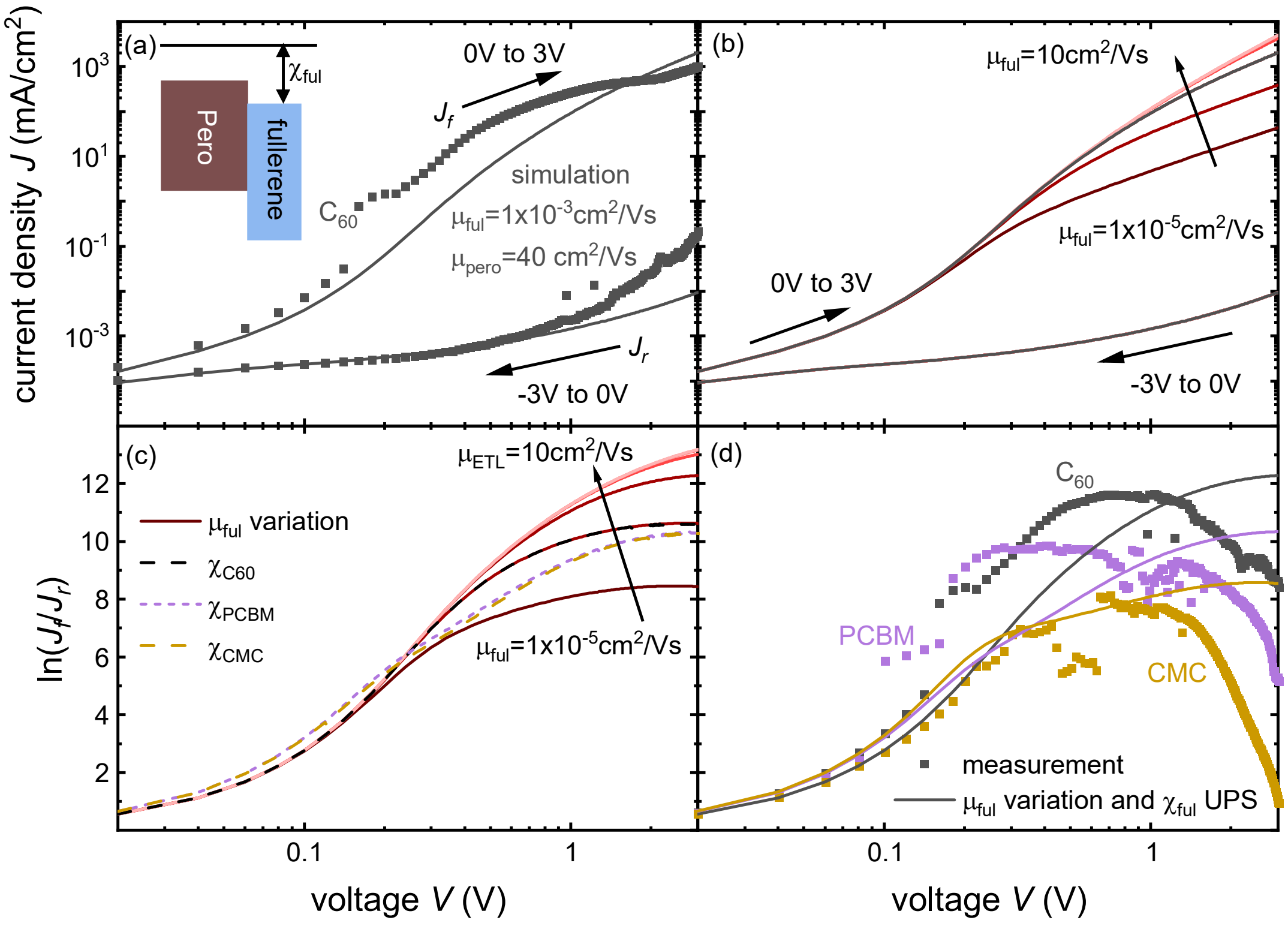

**Figure 2**: a) Dark *JV*-curves of the electron-only device with $C_{60}$ as ETL (squares) and the corresponding simulations using *ASA* (solid lines). The parameters can be found in Table S4. The lower branch with the arrow to the left is the measurement/simulation from -3V to 0V (reverse current $J_r$), the upper branch with the arrow to the right part determines the measurement/simulation from 0V to 3V (forward current $J_f$). b) *ASA* simulations for an electron-only device with different electron mobility in the fullerene ($1 \times 10^{-5}$ cm²/Vs to 10 cm²/Vs.). The curve does not change from -3V to 0V (lower branch) and increases for higher mobilities with increasing mobility in the fullerene. c) Logarithmic ratio of forward $J_f$ (0V to 3V) and reverse $J_r$ (-3V to 0V) current with the different fullerene mobilities used in b) (solid

lines) and with the same mobility but the electron affinity of the three fullerenes measured with UPS (dashed line). d) Logarithmic $J_f/J_r$ ratio calculated from the measurements of electron-only devices with different fullerenes (squares). The combination of different mobilities for each fullerene and the corresponding UPS electron affinity (lines) leads to $J_f/J_r$ ratios with a maximum similar to the measured curves.

**Figure 2a** shows the dark *JV*-curve of the electron-only device with $C_{60}$ as ETL (squares), measured from -3V to 3V on a double logarithmic plot. The lower branch is the current for negative (reverse) and the upper branch corresponds for positive (forward) voltages. The simplest method to derive the mobility from the curve would be the use of the Mott-Gurney equation S1 applied to the drift-limited region with slope close to two.[42] The linear fit to the region would lead to an apparent electron mobility $\mu_{\mathrm{app}} \approx 0.02\ \mathrm{cm}^2/\mathrm{Vs}$. However, the Mott-Gurney law is only valid under several conditions.[32] One condition is a strictly drift controlled current, where diffusion terms can be neglected. This condition is violated by asymmetric injection barriers and potentially by the existence of charged (and mobile) defects in the device. The asymmetric contacts can be clearly seen by the difference in current for negative and positive voltage scan.[36, 42] Furthermore, the Mott-Gurney law considers only one semiconductor sandwiched between two highly conductive contacts. Only the material between the contacts limits the electron mobility in the device. However, we used a device stack with perovskite and fullerenes that are known for their low conductivity[9]. To investigate the impact of low transport layer mobilities we performed simulations by varying the electron mobilities in fullerene and perovskite using the drift-diffusion simulation tool ASA.[43, 44] The simulated *JV*-curves for electron-only devices with a variable electron mobility in the fullerenes and in the perovskite are depicted in **Figure 2b** and **Figure S8**, respectively. The mobility $\mu_{\mathrm{ful}}$ varies from $1 \times 10^{-5}$ $\mathrm{cm}^2/\mathrm{Vs}$ to 10 $\mathrm{cm}^2/\mathrm{Vs}$. The lower branch does not change but the upper branch, i.e. the drift dominated regime increases with higher mobility in the fullerene and starts to saturate for mobilities between 1 $\mathrm{cm}^2/\mathrm{Vs}$ and 10 $\mathrm{cm}^2/\mathrm{Vs}$. This indicates that the current is limited by the lower mobility of the fullerene and perovskite layers. The red line in Figure 2a is the fitted *JV*-curve from the *ASA* simulations, where we used a mobility in the perovskite of $\mu_{\mathrm{pero}} = 29$ $\mathrm{cm}^2/\mathrm{Vs}$ while the mobility in the fullerene is $\mu_{\mathrm{C60}} = 9.5 \times 10^{-4}$ $\mathrm{cm}^2/\mathrm{Vs}$. The mobility for perovskite fits to previously measured mobilities in perovskite.[45-47] The fullerene mobility is low for $C_{60}$ but in the range of other fullerenes.[9, 48] Figure S7 shows the electron-only devices containing the three different fullerenes. All devices have asymmetric injection barriers, as the curve from -3V to 0V (arrow to the left) produces significantly less current than the measurement from 0V to 3V (arrow to the right).

In a single carrier device with one layer between two asymmetric contacts it is possible to calculate the built-in voltage [36] via

$$V_{\mathrm{bi}} = \frac{k_{\mathrm{B}}T}{q} \ln\left(\frac{8}{9} \frac{qN_{\mathrm{c}}d^2}{\varepsilon_{\mathrm{r}}\varepsilon_0} \frac{J_{\mathrm{f}}}{J_{\mathrm{r}}} \frac{1}{V}\right), \quad (1)$$

which is independent of the mobility of the layer and only depends on the thickness $d$, the density of states in the conduction band $N_C$, the dielectric constant $\varepsilon_r$ and the ratio of the forward $J_f$ (0V to 3V) and reverse $J_r$ (-3V to 0V) current. Following equation 1, the built-in voltage depends on material properties and the logarithmic ratio of the forward and reverse current. However, in Figure 2 b) we see that this ratio depends heavily on the mobility of the fullerene which is not factored in the equation. The solid lines in Figure 2c) show the logarithmic $J_f/J_r$ for the simulated curves with different fullerene mobility. Since we have two layers, i.e., perovskite and fullerene, the ratio does not only depend on the barrier created by the built-in voltage but also the barrier created by the offset between perovskite and fullerene. Therefore, Figure 2c) shows $J_f/J_r$ for the three different electron affinities given by the UPS measurement (dashed lines). The ratio is highest for the electron affinity of $C_{60}$ and nearly equal for PCBM and CMC. However, from Figure 2b, it is clear that the electron mobility of the fullerene has a larger influence on the $J_f/J_r$ ratio than the electron affinity. Figure 2 d) shows the current ratio calculated from the three electron-only devices with different fullerenes (squares) and calculated for simulations with the electron affinity from UPS and mobilities selected such that the maximum of the simulated ratio is close to the measurement results. Due to the large influence of electron mobility in the fullerene the ratio cannot be used to get an inside on the band alinement of the perovskite and the different fullerenes. However, the ratio between forward and reverse current may help to estimate the relevant mobility in the CTL.

## 2.3 Cell Performance

In the following, we present the effect of the different HTLs and ETLs on the performance of the perovskite solar cells. **Figures 3a** and **3b** shows the best (highest efficiency) performing *JV* -curves, in forward (dashed lines) and backward scan, of devices employing different combination of HTLs and ETLs, respectively. The *JV*-curve (black line) of Me-4PACz/ $Cs_{0.05}(FA_{0.83}MA_{0.17})_{0.95}Pb(I_{0.83}Br_{0.17})_3$ perovskite/$C_{60}$ based device is shown as a reference in both figures. All cells were stabilized with several *JV*-measurements under LED illumination (**Figure S12a**). As indicated by maximum power point tracking (MPPT) (**Figure S12b)**, for an example device ($V_{max}$ = 0.98 V) showing efficiency of 19.6% under one Sun illumination and 19.5% with MPP tracking, the cells are stable for several minutes. **Figure 3c-f** shows the box plots highlighting the influence of CTLs employed on the device characteristics of perovskite solar cells.

Devices employing different SAMs (Me-4PACz, 2PACz and MeO-2PACz) with $C_{60}$ ETL showed a similar device performance while Me-4PACz based cells showed slightly better efficiency owing to a higher *FF* compared to MeO-2PACz and 2PACz as HTL based cells. Nevertheless, devices with 2PACz HTL showed overall better $J_{SC}$ and $V_{OC}$ compared to Me-4PACz and MeO-2PACz case. PTAA based cells showed overall the lowest device performance with all device parameters ($J_{SC}$, $V_{OC}$ and *FF*) being lower than that of the solar cells employing SAMs. The PTAA cell with the highest efficiency (shown in Figure 3a) is the only device with such a high *FF*.

When changing ETLs while keeping Me-4PACz HTL constant, we observed that the device with $C_{60}$ as ETL showed the highest efficiency owing to high *FF* compared to the case with PCBM and CMC as ETLs. However, it was interesting to observe that by replacing $C_{60}$ with PCBM and CMC, the devices showed improvement in the $V_{OC}$. Thus, the ETL variation shows an anticorrelation between the recombination at open circuit and the recombination and resistive effects that happen during transport (i.e. voltages below $V_{OC}$). The key values for the best performing devices for all cases and the median cells are tabulated in **Table S5**.

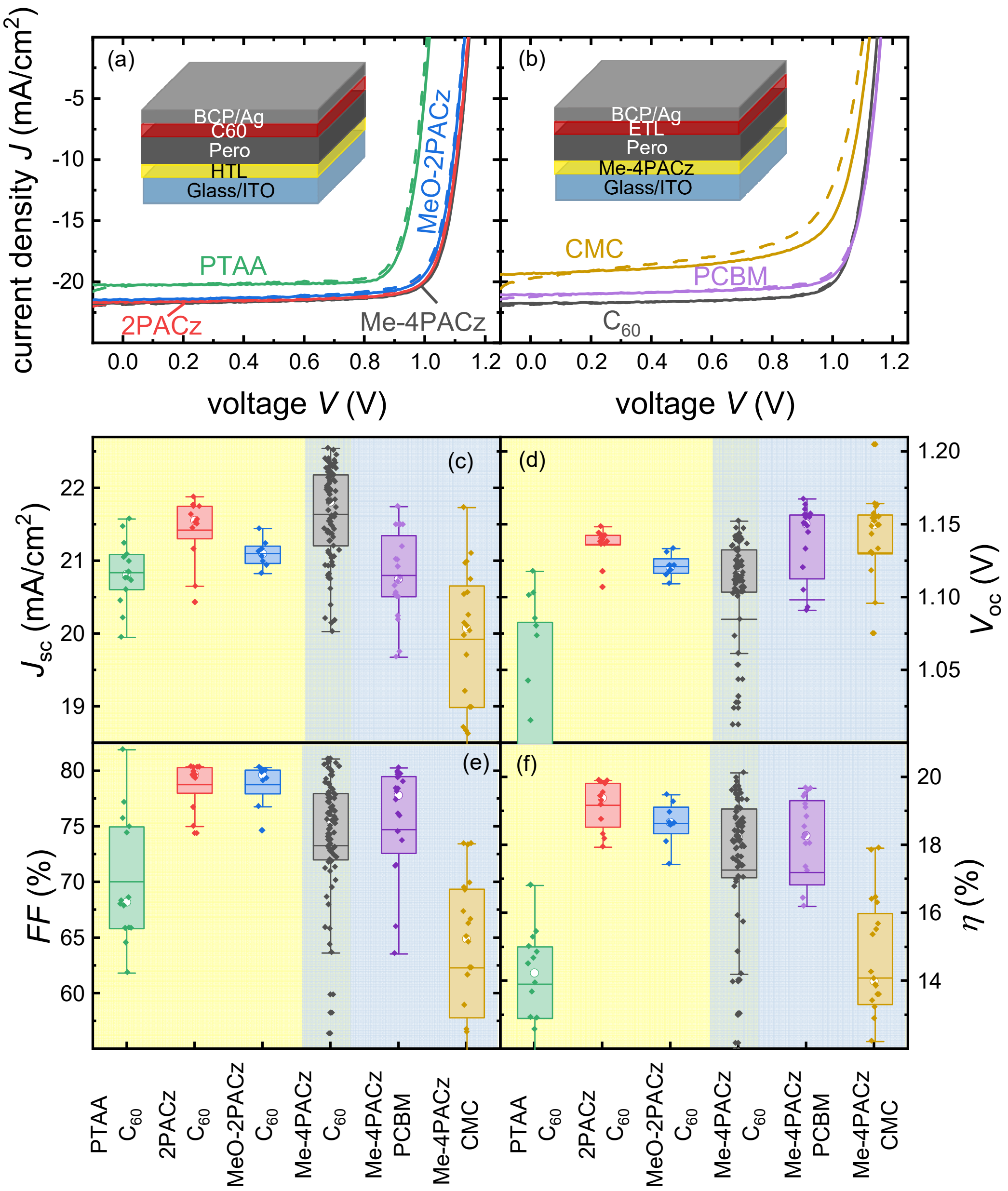

**Figure 3:** Illuminated *JV* -curves measured with a AAA Sun simulator in forward (dashed line) and backward (solid line) scan directions. (a) Variation in the HTLs with $C_{60}$ ETL as constant. (b) Variation in the ETLs by keeping Me-4PACz HTL as constant. (c)-(f) Statistic of (c) short circuit current $J_{SC}$, (d) open circuit voltage $V_{OC}$, (e) fill factor $FF$ and (f) efficiency $\eta$ for all the cases studied. The filled symbols are the measured data, the box contains 50% of the data, the bars give the mean values and the circles the median.

## 2.4 Charge transport

One selection criterion for CTLs is good charge extraction.[49] This means that the film should be as conductive as possible and should not present a barrier for extraction as discussed in section 2.2.

Therefore, one way to determine the quality of the CTLs is to compare the series resistances ($R_s$) of the cells comprising different CTLs. The $R_s$ is traditionally understood as an ohmic or nearly ohmic resistance in series with the diode (i.e. the non-linear, rectifying part of the solar cell). However, as resistive effects can also be non-ohmic (i.e. non-linear with voltage), it is often not possible to determine a single value for the $R_s$. Furthermore, the $R_s$ will always depend on the method used to determine its value as there is not one universal approach to quantify resistive losses. This section considers the $R_s$ as a function of voltage (or current) obtained by comparing dark and light *JV*-curves and discusses its influence on the *FF*.[9, 12] Furthermore, we consider charge extraction through the ETL by voltage-dependent PL measurements and compare the result with the initial decay of PL transients.

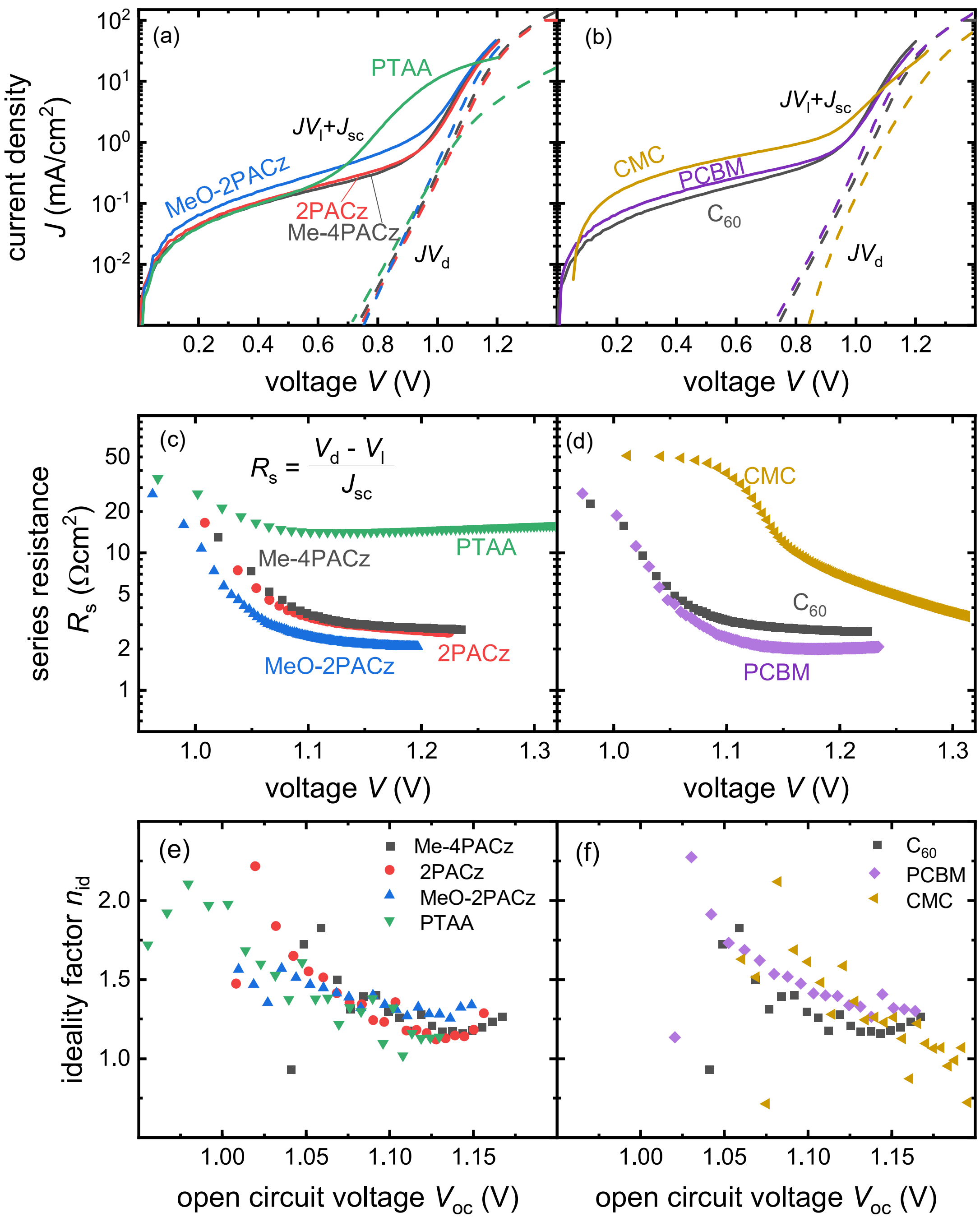

**Figure 4:** Dark (dashed line) and illuminated (solid line) $JV$-curves plotted in the first quadrant. a) devices with different HTLs and $C_{60}$ as ETL and b) devices with Me-4PACz as HTL and different ETLs. The Me-4PACz and $C_{60}$ curve are the same curve. $R_s$ calculated from the difference between dark and illuminated $JV$ -curves. $R_s$ calculated for devices c) with different HTLs and with $C_{60}$ as ETL and d) with different ETLs with Me-4PACz as HTL. e) and f) Ideality factor derived from the Suns-$V_{OC}$ measurement.

**Figures 4a** and **4b** shows the dark $JV$ -curves (dashed line) and the shifted illuminated $JV$-curves (solid line) on a semilogarithmic scale for devices with different HTLs and different ETLs respectively. In case of the illuminated $JV$-curves (Figure 4b), the region with a moderate slope at low voltages show a weakly voltage dependent recombination current, i.e., the recombination loss at $J_{SC}$. This so called photo-shunt is an indication for poor carrier extraction at low voltages most likely caused by the finite conductivity of the undoped transport layers.[12] A quantitative evaluation of the shunt is not easy, however, by comparing the currents at low voltages, as shown in Figure 4a, it becomes clear that there is no significant difference in charge transport for different HTLs. For the ETLs (Figure 4b), CMC has the highest photo-shunt indicating a high carrier concentration at short circuit and low forward bias caused by poor charge extraction. This finding is consistent with the reduced $FF$ and $J_{SC}$ (Figure 3) for CMC-based devices. However, the statistical fluctuations for CMC-based devices are rather high. The 50% range of $J_{SC}$ data starts from 19mA/cm$^2$ and ends at 20.7mA/cm$^2$ and the $FF$ ranges from 57.9% to 69.4%. Although there are cells that deviate more upwards or downwards.

The difference between the dark and illuminated $JV$-curves at higher voltages provides information on the $R_s$ of the cell. **Figure 4c** and **4d** shows the voltage dependent $R_s$ for the devices with different HTLs and ETLs respectively. For lower voltages ($V_d$ < 1.15 V), the $R_s$ is high mainly because of increasing internal $R_s$ in the cell while for higher voltages (from $V_d \geq$ 1.15 V), the curves show a flat region corresponding to the external $R_s$. The $R_s$ for the PTAA based cell (with $C_{60}$ as ETL) shows the highest external $R_s$ of 15.3 Ωcm$^2$ in a voltage range of $V_d$ = ~1.15 V to ~1.3 V while lowest $R_s$ of 2.0 Ωcm$^2$ was observed in the same voltage range for the device with MeO-2PACz as HTL (with $C_{60}$ as ETL). Furthermore, among various ETLs, the PCBM based cell (with Me-4PACz SAM as HTL) showed lowest $R_s$ of 2.0 Ωcm$^2$. This suggests that both the ETL and the HTL are limiting the $R_s$ of the cell and the choice of suitable combination of CTLs can help to reduce the $R_s$ and improve the $FF$ of the cell. To further confirm this conclusion, we fabricated device with MeO-2PACz and PCBM as HTL and ETL, respectively, and the resultant perovskite device showed one of the highest efficiency $\eta$ of 21% with $FF$ of ~83 %. The device $JV$-curve is shown in **Figure S14**. **Figures 4e** and **4f** shows the voltage dependent ideality factor for the cells with different CTLs. The ideality factor at high voltages is constant for a small range of voltages from 1.1V to 1.15V. In this region we calculate the cell ideality factor by the arithmetic mean. For CMC two regions with nearly constant ideality factor can be found, from 1.2V to 1.15V and from 1.15V to 1.1V, the ideality factor is $n_{id}$=1.06 and $n_{id}$=1.37. For the other cells the ideality factors for the cells lie between $n_{id}$=1.13 and $n_{id}$=1.28. For the different HTLs, PTAA shows the lowest $n_{id}$ while MeO-2PACz shows the highest. For different ETLs, the $n_{id}$ of $C_{60}$ is lower than that of PCBM and both are in between the two ideality factors of CMC. All ideality factors are close to $n_{id}$=1 suggesting that recombination dominantly happens in regions of the cell, where the electron and hole densities are unequal or the position of the most recombination-active defect within the band gap is far away from mid gap (shallow defect).

To further investigate the influence of series resistance and ideality factor on the cell performance, we compare the power density of the illuminated *JV* curve (blue) and the reconstructed power density from $J_{SC}/V_{OC}$ (red) (**Figure 5**) (see SI paragraph 2.4.2). By showing the power-density voltage curve rather than the *JV* curve, the efficiency of the cell can be easily read from the maximum of the curves. The difference in efficiency and *FF* between illuminated *JV* curve and $J_{SC}/V_{OC}$ curves is due to the absence of voltage losses over series resistances in the $J_{SC}/V_{OC}$ measurement ($\Delta FF_{R_s} = FF_{Jsc/Voc} - FF_l$). In addition, we simulated the ideal curve using the measured $J_{SC}$ and radiative $V_{OC}$ (green). By comparing these ideal curves with the ones constructed from the $J_{SC}/V_{OC}$ data ($\Delta FF_{n_{id}} = FF_{ideal} - FF_{Jsc/Voc}$), we can determine the loss due to non-ideal diode behaviour, i.e. $n_{id} \neq 1$. From Figure 4, we can see that the loss due to the series resistance is highest for PTAA ($\Delta FF_{R_s} = 19.0\%$ ) and CMC ($\Delta FF_{R_s} = 20.4\%$). In addition to this, Figure 4 suggests the loss due to $R_s$ is low for the combination with PCBM ($\Delta FF_{R_s} = 5.3\%$) and MeO-2PACz ($\Delta FF_{R_s} = 6.2\%$), However, despite the higher $R_s$, the lowest loss occurs for Me-4PACz ($\Delta FF_{R_s} = 4.8\%$). Next, we look at the loss due to $n_{id}>1$. We observe the lowest loss for 2PACz ($\Delta FF_{n_{id}} = 2.3\%$), the loss for CMC is low as well ($\Delta FF_{n_{id}} = 2.9\%$), which is consistent with the high $V_{OC}$ of the cell and the ideality factor close to $n_{id} = 1$. For all combinations the *FF* loss arises either in equal parts from $R_s$ and non-ideal diode behaviour or predominantly because of the series resistance (especially for PTAA and CMC). All values are tabulated in Table 1. Figure S18 shows the correlation between the series resistances and the ideality factors with the respective losses in *FF*. It is noticeable that the accuracy of $n_{id}$ determination is not very good, since no clear plateau can be recognized. Therefore, no clear trend between the data can be recognized.

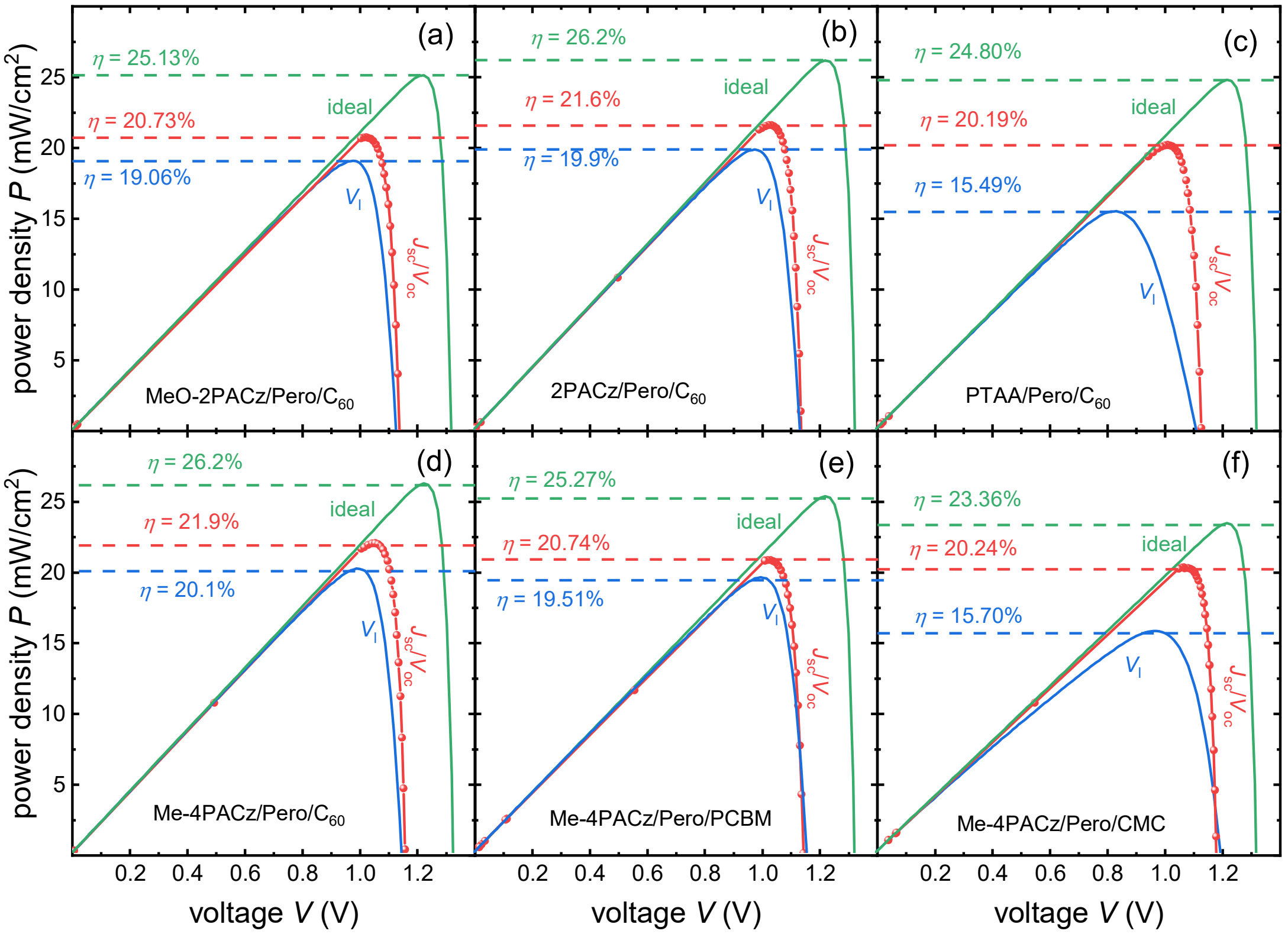

**Figure 5:** Power density of the illuminated cell (blue), reconstructed form $J_{SC}/V_{OC}$ curve (red) and ideal curve (green) for cells with different (a – c) HTLs and (d – f) ETLs. The dashed lines give the maximum power density, i.e. the efficiency of the measured, series resistance free and ideal solar cell.

**Table 1:** *FF* difference between solar cells with different HTLs and ETLs and $R_s$ free *FF* from $J_{SC}/V_{OC}$ curve $\Delta FF_{R_s}$ and difference between ideal *FF* and from $J_{SC}/V_{OC}$ *FF*, $\Delta FF_{n_{id}}$

| **Sample** | $\Delta FF_{R_s}$ **(%)** | $\Delta FF_{n_{id}}$ **(%)** |
|---|---|---|
| MeO-2PACz/Pero/$C_{60}$ | 6.2 | 4.0 |
| 2PACz/Pero/$C_{60}$ | 7.9 | 2.3 |
| PTAA/Pero/$C_{60}$ | 19.0 | 4.3 |
| Me-4PACz/Pero/$C_{60}$ | 4.8 | 5.0 |
| Me-4PACz/Pero/PCBM | 5.3 | 5.3 |
| Me-4PACz/Pero/CMC | 20.4 | 2.9 |

To study the losses in $J_{SC}$ due to non-ideal extraction of the photogenerated charge carriers by the CTLs, we measure the voltage dependent photoluminescence PL(V)[23-26, 50] for three perovskite solar cells of different ETLs ($C_{60}$, PCBM, and CMC). From the PL(V) measurement with the obtained $J_{SC}$ and the ideality factor $n_{id}$ for each device, the recombination current density ($J_{rec}$) can be evaluated from[12]

$$J_{\mathrm{rec}} = \frac{J_{\mathrm{sc}}\phi(V)^{\frac{1}{n_{\mathrm{id}}}}}{\phi_{\mathrm{oc}}^{\frac{1}{n_{\mathrm{id}}}} - \phi_{\mathrm{sc}}^{\frac{1}{n_{\mathrm{id}}}}} \tag{2}$$

where $\phi_{\mathrm{oc}}$ is the luminescence at open circuit and $\phi_{\mathrm{sc}}$ the luminescence at short circuit.

**Figure 6a** shows the quasi-Fermi level splitting ($\Delta E_{\mathrm{F}}$) as a function of external voltage for three cells with $C_{60}$, PCBM and CMC that we extracted from the photoluminescence intensity. Note that these cells are not identical to the ones discussed in Figure 4 and 5 but were made with the same recipes. If carrier extraction is efficient, the $\Delta E_{\mathrm{F}}$ at short circuit should be as small as possible. At 0V, $C_{60}$ shows the smallest $\Delta E_{\mathrm{F}}$=1.03eV, CMC has an intermediate $\Delta E_{\mathrm{F}}$=1.07eV while PCBM shows the highest $\Delta E_{\mathrm{F}}$=1.10eV. **Figure 6b** shows the respective voltage-dependent recombination currents of the three ETLs versus the applied bias. For voltages up to 0.9 V, the recombination current remains constant and then increases strongly with the voltage. In the case of a short circuit, $C_{60}$ shows the smallest recombination current density ($J_{\mathrm{rec}}$ (0V) = 1.25 mA/cm$^2$), and for CMC $J_{\mathrm{rec}}$ (0V) = 1.72 mA/cm$^2$, however, PCBM shows a slightly larger recombination current density ($J_{\mathrm{rec}}$ (0V) = 2.55 mA/cm$^2$). Thus, the extraction of the photogenerated charges in $C_{60}$ is the best while PCBM leads to the worst collection. This stands in contrast to the lower series resistance measured for cells with PCBM. The illuminated *JV*-curves for the samples we measured in Figures 4 and 6 are shown in Figure S19. The CMC sample measured in Figure 4 has a lower $FF_{\mathrm{Rs}}$=67% compared to the cell measured for Figure 6 $FF_{\mathrm{PL}}$=77%, the cells with other ETLs behave similar for both measurements. The series resistance, ideality factor and the equivalent of Figure 5f are shown in Figure S20.

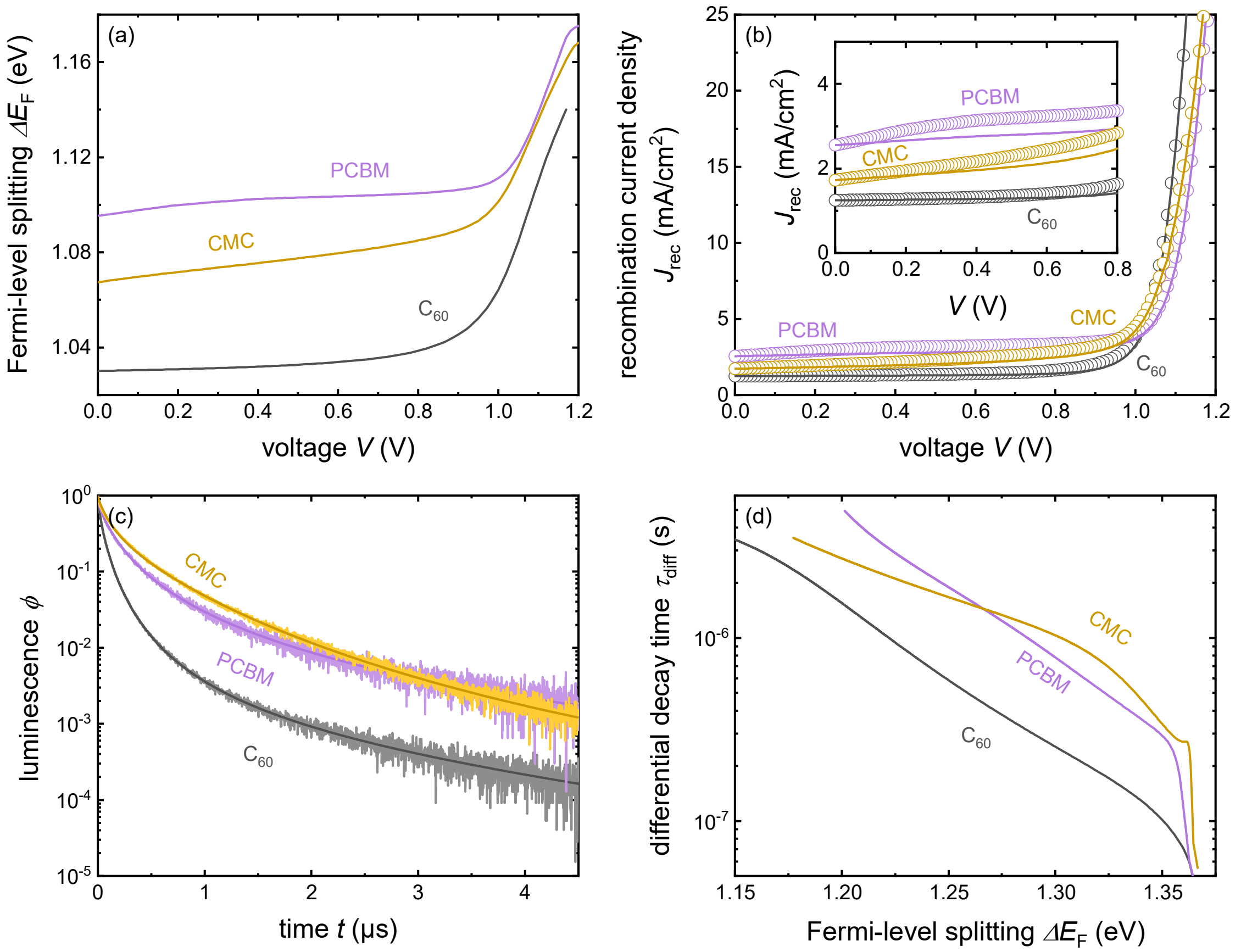

**Figure 6:** (a) The Fermi-level splitting versus the external bias obtained from the voltage-dependent photoluminescence measurement for three perovskite solar cells using either $C_{60}$, PCBM, or CMC as ETL. (b) The recombination current density as a function of the external voltage that was calculated from the PL(V) data using equation (2)[12] (open circles) with an ideality factor of 1.2 for $C_{60}$, 1.33 for PCBM, and 1.36 for CMC, or from the light *JV*-curve that was measured using the PL(V) setup and the maximum possible photocurrent for each device (solid lines) such that $J_{\text{rec}} = J_{\text{light}} + J_{\text{ph,max}}$. A zoom in the region of SC and low forward bias is shown, $J_{\text{rec}}$ at SC fits perfectly from both the PL(V) and the up shift of the light *JV*-curve by $J_{\text{sc}}$ for the three devices. (c) Tr-PL-decay measured with 258.2 nJ/cm$^2$ laser intensity of the layer stack Me-4PACz/Pero/ETL for varying ETLs. A rational function was fitted to the background-corrected decay data of each curve. (d) Differential decay times calculated from the fitted tr-PL data as a function of Fermi-level splitting.

In addition to the voltage dependent PL, the transient PL was measured on the HTL/perovskite/ETL device stack layers without BCP/Ag. **Figure 6c** shows the normalized PL decay versus time. Although the interpretation of tr-PL curves in multi-layer systems is sufficiently complicated, since both transport and recombination are measured, we consider the tr-PL decay in the context of carrier extraction. Without a quencher, a rapid carrier loss is a sign of strong recombination, but when a transport layer is added, a rapid decrease in PL can also mean that the carriers are extracted efficiently.[27] Unlike the PL(V), no voltage can be applied to the layers, so the half cells are always under open circuit condition. This means that transport and recombination processes always coexist and cannot easily be disentangled.

Figure 6c shows a faster decay for $C_{60}$ and a slower decay for PCBM and CMC with PCBM decreasing faster at first but then slowing down while CMC decreases faster. Since the observation is notably consistent with the good extraction for $C_{60}$ from the PL(V) data, we assume that the tr-PL decay in this case is dominated by the transfer of electrons and holes to the ETL and HTL layers.

We then calculated the differential decay time

$$\tau_{\text{diff}} = -\frac{n(t)}{dn(t)/dt} = -\frac{2\phi(t)}{d\phi(t)/dt} \tag{3}$$

which assumes that the perovskite layer is well approximated as an intrinsic semiconductor, i.e. $\phi \sim np = n^2$ for the electron ($n$) and hole densities ($p$) encountered during the PL transient. **Figure 6d** shows the differential decay time from the curves fitted to the data in Figure 6c. The derivative is no longer plotted against time but against the Fermi-level splitting, as described by ref [51]. A feature that stands out in Figure 6d is that the differential decay time is continuously changing with $\Delta E_F$ and not giving a constant value that could be considered a recombination lifetime. To further investigate the recombination processes in the cell we looked at the tr-PL of the bare perovskite and perovskite with $C_{60}$ on all four HTLs, and Me-4PACz/Pero with different ETLs and the details are explained in the next section.

## 2.5 Non-radiative recombination losses.

For perovskite solar cells nonradiative recombination at the perovskite CTL interfaces plays a critical role for the recombination processes and thus the $V_{OC}$ of the device.[29] In order to characterize bulk and interface recombination losses, typically both steady-state and transient PL measurements are applied to a variety of layer stacks (from the film to the full device). From the differences between the different samples, conclusions are derived about the recombination activity of the bulk and the different interfaces. As this approach has been applied to both steady-state and transient PL,[51] the complexity of data interpretation differs hugely between the two methods. In steady-state PL, applied to a series of samples with the same absorber but different interfaces, the absolute PL intensity corrected for variations in outcoupling between the samples will clearly identify differences in the rates of non-radiative recombination. [52, 53] Lower PL intensities will always identify situations, where the contribution of non-radiative recombination has increased. It is therefore deceptive to use steady-state PL quenching at open circuit as a positive signal of charge extraction as it only probes the negative implications of having a junction between an absorber and contact layer.[54] In transient PL, the situation is somewhat different as currents can be flowing during the transient even in uncontacted films and layer stacks that do not allow current flow through an external circuit. Thus, dynamic effects such as transfer of charge to contact layers can lead to faster decays as seen in Figure 6d and superimpose the effects of non-radiative recombination. Thus, characterizing recombination dynamics in multilayer systems with interfaces and contact layers is challenging due to the superposition of transport, recombination and capacitive effects that result from the charging or discharging of electrodes[55], charge transport[27] layers or shallow defects[56].

In the following, we will compare both transient and steady state PL and present a new approach to quantitatively disentangle contributions of recombination to the transient PL from contributions of transport and capacitive effects. For this purpose, we investigated three different layer stacks: ITO/HTL/Perovskite with the different HTLs, ITO/HTL/Perovskite/$C_{60}$ and ITO/Me-4PACz/Perovskite/ETL with the different ETLs. Figure 6c and 6d show the tr-PL data and the derived decay time for one illumination intensity. To increase the range of the quasi-Fermi level splitting, we performed the measurement under three different laser intensities (2598 nJ/cm$^2$, 258.2 nJ/cm$^2$ and 25.4 nJ/cm$^2$). Figure S23 a)-c) shows the composite tr-PL decay for one example of each layer composition. Figure S23 d)-f) then shows the derived differential decay times for all layer compositions. For layer stacks without ETL, a clear S-shape can be seen for all curves. We have previously identified pronounced S-shapes in simulations as originating from situations, where substantial densities of charge carriers are re-injected into the pool of free carriers in the perovskite absorber at later times[51]. This re-injection can originate from HTL or ETL or be caused by detrapping from defects that have a high density but only interact efficiently with one band. Figure S25 shows simulations with shallow defects that are filled during the early stages of the transient and then empty during later stages and thereby cause an S-shaped decay time curve. For the layers with ETL, the S-shape is less pronounced and the decay times at a given $\Delta E_F$ are slightly shorter as compared to samples without ETL. At lower quasi-Fermi level splitting (later time), the differential decay time increases continuously until it runs into the limit determined by the lowest repetition rate ($f_{\mathrm{rep}}$ = 20 kHz) of our system.

In order to disentangle recombination effects from transport and capacitive effects, one may express the differential equation for the electron density $n$ in the perovskite absorber as

$$\frac{dn}{dt} = -k_{\mathrm{rad}}(np - n_i^2) - \beta_{\mathrm{n}}\, n(N_{\mathrm{t}} - n_{\mathrm{t}}) + e_{\mathrm{n}}\, n_{\mathrm{t}} \tag{4}$$

the hole density $p$ as

$$\frac{dp}{dt} = -k_{\mathrm{rad}}(np - n_i^2) - \beta_{\mathrm{p}}\, pn_{\mathrm{t}} + e_{\mathrm{p}}\, (N_{\mathrm{t}} - n_{\mathrm{t}}) \tag{5}$$

and the density of a trap state $n_t$ as

$$\frac{dn_{\mathrm{t}}}{dt} = \beta_{\mathrm{n}}\, n(N_{\mathrm{t}} - n_{\mathrm{t}}) - \beta_{\mathrm{p}}\, pn_{\mathrm{t}} - e_{\mathrm{n}}\, n_{\mathrm{t}} + e_{\mathrm{p}}\, (N_{\mathrm{t}} - n_{\mathrm{t}}). \tag{6}$$

Equations (4) to (6) assume an absorber layer with a single (relevant) trap state that captures electrons and holes with the capture coefficients $\beta_{\mathrm{n,p}}$ and emits them back to the bands with the emission coefficients $e_{\mathrm{n,p}}$ that depend on the trap depth $E_t$ via $e_n = \beta_n N_c \exp\left(\frac{E_t - E_c}{kT}\right)$, and $e_p = \beta_p N_v \exp\left(\frac{E_v - E_t}{kT}\right)$.

If the density of trapped electrons hardly changes with time (i.e. if $dn_t/dt \approx 0$), $dn(t)/dt \approx -\bar{R}$ will be a good approximation. Here, the steady-state recombination rate $\bar{R} = \bar{R}_{\mathrm{rad}} + \bar{R}_{\mathrm{SRH}}$, whereby the terms on the right-hand side are the steady-state recombination rate for radiative and SRH recombination. This situation ($dn(t)/dt \approx -\bar{R}$) is likely to be relevant for instance if we have a low density of deep defects. Alternatively, if a defect becomes sufficiently shallow and its density becomes sufficiently high, trapping and detrapping rates become significant and $dn/dt \neq -\bar{R}$. Similarly, if we have a layer stack or a complete solar cell, exchange of electrons and holes with other layers or the electrodes can lead to $dn/dt \neq -\bar{R}$.

All effects leading to $dn/dt \neq -\bar{R}$ are capacitive insofar as they originate from charge carriers being temporarily stored in reservoirs (traps, other layers, electrodes) and then later being reinjected into the one charge-carrier reservoir that contributes to luminescence emission (the perovskite layer). Depending on the type of sample (film, layer stack, full device) some of these effects may be absent. In a steady-state PL measurement at open circuit, we know that the average rates of recombination ($\bar{R}$) and generation ($G$) must be identical. Thus, we could determine an effective lifetime $\overline{\tau_{\mathrm{eff}}}$ from the steady state PL measured at a light intensity leading to an average generation $G$ that would be given by $\overline{\tau_{\mathrm{eff}}} = \bar{n}/\bar{G} = \bar{n}/\bar{R}$. Furthermore, if recombination is dominant in the transient PL as well, the differential decay time obtained from eq. 3 is given by $\tau_{\mathrm{diff}} = \overline{\tau_{\mathrm{eff}}}$. To identify the relative contributions of recombination vs. capacitive effects to the decay time, we plot $\tau_{\mathrm{diff}}$ and $\overline{\tau_{\mathrm{eff}}}$ into the same figure and compare their relative values as a function of either $n$ or $\Delta E_{\mathrm{F}}$. As these would be related by $n = n_{\mathrm{i}}\exp(\Delta E_{\mathrm{F}}/2kT)$, there would be no fundamental difference between the two options.

Analogously, as we can calculate the decay times for recombination from steady state data, we could also do the inverse operation if the assumption $dn/dt = -\bar{R}$ was true. The steady state PL data is often plotted as $\Delta E_{\mathrm{F}}$ versus generation rate or photon flux expressed in suns. This way of plotting the data allows one to directly see what the upper limit for $V_{\mathrm{OC}}$ would be that a layer (stack) without contacts would impose on any device that includes these layers and interfaces. If $dn(t)/dt \approx -\bar{R}$ is a good approximation, we can assign an effective (steady-state) recombination or generation rate to the decay time via $G = n/\tau_{\mathrm{diff}}$ and determine the associated Fermi level splitting via $\Delta E_{\mathrm{F}} = 2kT\ln(n/n_{\mathrm{i}})$.

In the following, we will illustrate these different mathematical operations for a perovskite layer on glass, a layer stack including the HTL and one including both HTL and ETL. **Figure 7a** shows $\tau_{\mathrm{diff}}$ for three different laser intensities (solid line) determined by equation 3 as a function of $\Delta E_{\mathrm{F}}$. Early times in the transient correspond to the data at high $\Delta E_{\mathrm{F}}$, i.e. it is intuitive to discuss this type of plot from the (lower) right to the (upper) left. Each branch of $\tau_{\mathrm{diff}}$ contains two clearly distinguishable parts. At early times and high $\Delta E_{\mathrm{F}}$, the decay time increases rapidly until it approaches the dashed blue line. At this point it changes its slope and increases more slowly than before towards lower values of $\Delta E_{\mathrm{F}}$.

Independent of the laser fluence and the initial carrier density, all decay times eventually approach the same slope at the second part of each decay time vs. $\Delta E_F$ curve. The symbols representing the decay time $\overline{\tau_{\text{eff}}}$ calculated from the steady state data is slightly higher but still reasonably close to the blue dashed line. The dashed blue line has a constant slope that represents an ideality factor. For $n = p$, the slope would then follow from

$$\overline{\tau_{\text{eff}}} = \frac{\bar{n}}{R(\bar{n})} = \exp\left(\frac{\Delta E_{\text{F}}}{kT}\left[\frac{1}{2} - \frac{1}{n_{\text{id}}}\right]\right). \tag{7}$$

Thus, a deep trap in high-level injection would cause $n_{\text{id}} = 2$ and hence $\overline{\tau_{\text{eff}}}$ would be constant as a function of $\Delta E_{\text{F}}$.[51, 57] A recombination mechanism that causes $n_{\text{id}}$ to become smaller than 2 (e.g. shallow trap or radiative recombination) would lead to an increase of $\overline{\tau_{\text{eff}}}$ towards lower values of $\Delta E_{\text{F}}$. We note that the ideality factor ($n_{\text{id}} = 1.6$) that originates from the steady state PL via

$$n_{\text{id}} = \frac{d\Delta E_{\text{F}}}{kTd\ln(G)}. \tag{8}$$

is roughly identical to the ideality factor that is consistent with the slope of the tr-PL in Figure 7a. Thus, both the slope of $\tau_{\text{diff}}$ vs. $\Delta E_{\text{F}}$ as well as the absolute value suggest that the decay times of the film on glass are primarily affected by recombination and not by trapping/detrapping effects at least in the part of the decay that is close to the blue dashed line in Figure 7a. At short times and high $\Delta E_{\text{F}}$, $\tau_{\text{diff}}$ increases rapidly with decreasing $\Delta E_{\text{F}}$ which constitutes a clear deviation from the approximation $dn(t)/dt \approx -R$. In a perovskite film, this deviation can be explained either by charge-carrier diffusion or charge-carrier trapping. In case of diffusion, the logic would be that the carrier density is not only a function of $t$ but also a function of the spatial position $x$ within the film. Therefore, the equation to solve would be a partial differential equation of the form $dn(t,x)/dt = -R(x,t) + D_{\text{n}}\, d^2n/dx^2$, where $D_{\text{n}}$ is the diffusion constant. The laser pulse creates more charge carriers close to the front surface than further away from it. Because $\phi \sim np = n^2$, diffusion will lead to a homogenization of the carrier concentration and in consequence to a reduction in PL intensity even if the average density of charge carriers doesn't change.[58, 59][60, 61] In case of carrier trapping, initially empty traps would be filled by free electrons or holes leading to a reduction in PL that is faster than the rate of recombination.

**Figures 7b** and **7c** show the decay times calculated for perovskite films on Me-4PACz and Me-4PACz/perovskite/$C_{60}$, respectively. Interestingly, the decay times are higher than for the film on glass and increase more steeply towards lower $\Delta E_{\text{F}}$ which is consistent with a lower ideality factor (see blue dashed lines). In addition, we now observe significant differences between $\tau_{\text{diff}}$ and $\overline{\tau_{\text{eff}}}$. The decay time from tr-PL becomes longer while the decay time from the steady state PL only slightly increases relative to the film on glass. Thus, we have to be in a range, where the reservoir of free charge carriers in the perovskite is refilled to a significant extent at later times during the decay. Thus, either de-trapping or reinjection from the HTL must have a strong effect on the transient data but no effect on the steady state data. In the presence of the $C_{60}$ ETL, the shift between steady state and transient PL decay times becomes even more pronounced. In the SI, section 2.5.4, we visualize the effect of de-trapping on the decay times from transient PL simulation shows that for a trap near the conduction band the lifetime from ss-PL is indeed smaller than the tr-PL lifetime (Figure S30).

**Figures 7d** to **7f** mirror the information content of Figures 7a to 7c by presenting each dataset from the above panel in the logic of steady state PL. The squares represent the actual steady-state PL measurement while the lines represent the suns-$\Delta E_F$ data that is reconstructed from the transient PL under the assumption $dn(t)/dt \approx -\bar{R}$. The information content of Figures 6d to 6f is identical to Figures 6a to 6c and shows that the values of $\Delta E_F$ that one would derive from transient PL are substantially higher than the values from the steady state PL and are therefore not correct. They only serve to visualize the fact that the condition $dn(t)/dt \approx -\bar{R}$ is not met in the samples with HTL or ETL.

**Figure 7g** shows the quasi-Fermi level splitting at one Sun for ss-PL and tr-PL as well as the median $V_{OC}$ of a cell with Me-4PACz and $C_{60}$ as transport layers. As mentioned before, for perovskites on glass there is no significant difference in $\Delta E_F$ from ss-PL and tr-PL, the highest potential of $\Delta E_F$ and thus the largest loss to the real cell has the layer on Me-4PACz without $C_{60}$. The $C_{60}$ reduces the $\Delta E_F$, but when the traditional $\Delta E_F$ from ss-PL is considered, the loss is not as big as in the tr-PL case. The comparison of the $\Delta E_F$ from ss-PL and tr-PL with the respective median $V_{OC}$ for the different HTL and ETL combinations is shown in **Figure 7h** and Figure S29 respectively. The difference between $\Delta E_{F,\text{ss-PL}}$ and $\Delta E_{F,\text{tr-PL}}$ is largest for systems with $C_{60}$ and decreases for PCBM and more even for CMC. The largest $V_{OC}$ potential, i.e. the largest $\Delta E_F$, is also found for CMC. However, the median $V_{OC}$ of cells with CMC is not significantly larger than the median $V_{OC}$ of cells with PCBM. Thus, the loss from half cell to cell is greatest for CMC. Also, in the system with PTAA, a large loss between $\Delta E_F$ and $V_{OC}$ is observed.

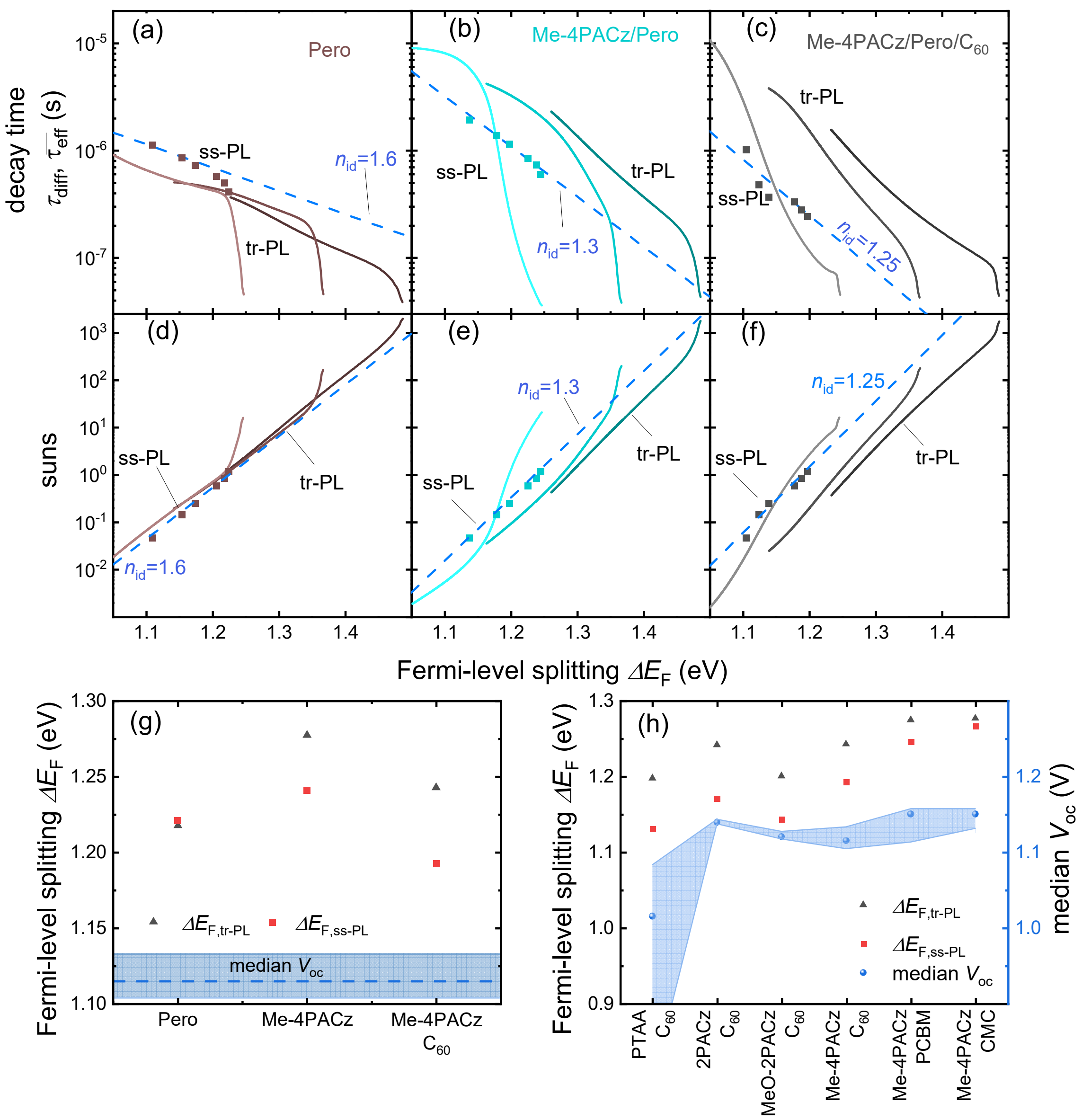

**Figure 7**: a)-c) Decay time as derived from tr-PL measurements $\tau_{\text{diff}}$ (solid line) and steady state PL $\overline{\tau_{\text{eff}}}$ at different illumination intensities (squares) and computed with eq. 7 for the best fitting ideality factor to the steady state data (dashed line). a) Perovskite on glass, b) perovskite on Me-4PACz and c) half cell with Me-4PACz/Pero/$C_{60}$. d)-f) Illumination intensity in Suns vs. Fermi level splitting for ss-PL (squares) and tr-PL (lines) as well the calculated from eq. 8 (dashed line). For the same stacks as a)-c). g) quasi fermi level splitting calculated from ss-PL (red squares) and tr-PL (black triangles) at one sun for perovskite on glass, on Me-4PACz and on Me-4PACz covered with $C_{60}$. The median $V_{\text{oc}}$ of the cell with the last configuration is shown as dashed line with the blue region representing 50% of the data as shown by the box in Figure 3. h) fermi-level splitting of the different cell stacks from ss-PL (red squares) and tr-PL (black triangles) as well as median $V_{\text{oc}}$ (blue dots) of the respective cells.

## 3. Conclusion

The right choice of CTL is important for obtaining high efficiencies in perovskite solar cells as it affects both recombination as well as efficient extraction of charge carriers. In order to minimize recombination and maximize the efficiency of extraction, any CTL has to fulfil a range of requirements such as the right energy levels, good conductivity and low interfacial recombination. In this paper, we have presented different methods to identify and quantify the properties that constitute a suitable HTL and ETL. Initially, we investigated the question of band alignment, where we found that the most severe obstacle for quantitative statements about the quality of band alignment is the uncertainty in the band edges of the perovskite relative to vacuum. There are three different approaches to determine the valence band edge of the perovskite when measuring the energy level with UPS.[18-20] The combination with simulations showed that a value between the linear and logarithmic method reflects the reality in the cell best. Nevertheless, no exact band alignment between the transport layers and the perovskite can be determined. Therefore, other methods are needed to quantify the offset between perovskite and CTL. Here, we have looked at electron-only devices with perovskite and fullerene and tried to determine the barrier at the interface via the ratio of reverse to forward current. However, drift-diffusion simulations have shown that unlike simple (perovskite only or fullerene only) devices, the formula for the built-in voltage[36] does not apply, but the mobility of the fullerene significantly affects the ratio between $J_r$ and $J_f$. In addition to band alignment, we studied charge transport in the cells by quantifying the series resistance for different HTLs and ETLs. The lowest series resistances were found for cells with MeO-2PACz/Pero/$C_{60}$ and for Me-4PACz/Pero/PCBM. A closer look at the losses in the *FF* shows that the CTL-to-CTL variations in *FF* losses are dominated by variations in $R_s$ rather than variations in the ideality factor. Voltage-dependent and transient PL measurements show that charge extraction is most efficient in solar cells with $C_{60}$ while samples with PCBM show significantly less efficient charge carrier extraction.

To better understand non-radiative recombination, steady state PL and transient PL were investigated in combination. Layers without ETL show an S-shape which could be reproduced in simulations by defects near the band edge. By comparing the Fermi-level splitting in a sum of tr-PL and ss-PL, it could be shown that for layer systems without ETL or with $C_{60}$, the Fermi-level splitting determined from the tr-PL is clearly larger than that determined classically with ss-PL. In other words, the steady state lifetime is shorter than the differential decay time from the tr-PL. This indicates an additional dynamic process that takes place at a later time, such as a detrapping effect from flat defects or reinjection of the charge carriers from $C_{60}$ or one of the HTLs.

### Acknowledgments

The authors acknowledge support from the Helmholtz Association via the Helmholtz Young-Investigator Group Frontrunner and via the project-oriented funding (POF IV). We also acknowledge funding from the DFG for the project CREATIVE within the SPP "Perovskite Semiconductors: From Fundamental Properties to Devices" (SPP 2196). A.K. thanks the Helmholtz Young Investigator Group FRONTRUNNER. M.S. acknowledges funding by ProperPhotoMile. Project ProperPhotoMile is supported under the umbrella of SOLAR-ERA.NET Cofund 2 by The Spanish Ministry of Science and Education and the AEI under the project PCI2020-112185 and CDTI project number IDI-20210171; the Federal Ministry for Economic Affairs and Energy on the basis of a decision by the German Bundestag project number FKZ 03EE1070B and FKZ 03EE1070A and the Israel Ministry of Energy with project number 220-11-031. SOLAR-ERA.NET is supported by the European Commission within the EU Framework Programme for Research and Innovation HORIZON 2020 (Cofund ERA-NET Action, No. 786483). Funded by the European Union. Views and opinions expressed are however those of the author(s) only and do not necessarily reflect those of the European Union or European Research Council Executive Agency (ERCEA). Neither the European Union nor the granting authority can be held responsible for them. The authors acknowledge funding from the European Research Council under the Horizon Europe programme (LOCAL-HEAT, grant agreement no. 101041809).
J.S. and A.K. contributed equally to this work.